\documentclass[12pt]{article}
\usepackage[]{epsfig}
\usepackage{amsfonts}
\usepackage{amsmath}
\usepackage{color}

\newcommand{\be}{\begin{equation}}
\newcommand{\ee}{\end{equation}}
\newcommand{\ba}{\begin{eqnarray}}
\newcommand{\ea}{\end{eqnarray}}

 \newcommand{\bea}{\begin{eqnarray}} \newcommand{\eea}{\end{eqnarray}}

\def\vr{{\check e}_r}
 \def\vteta{{\check e}_\theta}
 \def\vfi{{\check e}_\varphi}

\begin{document}
\title{\bf Holography and $AdS_4$ self-gravitating dyons}
\author{A.~R.~Lugo$^a$, E.~F.~Moreno$^b$  and
F.~A.~Schaposnik$^a$\thanks{Associated with CICBA}
\\
{\normalsize $^a\!$\it Departamento de F\'\i sica, Universidad
Nacional de La Plata}\\ {\normalsize\it C.C. 67, 1900 La Plata,
Argentina}
\\
{\normalsize $^b\!$\it Physics Department, Northeastern University}\\
{\normalsize\it Boston, MA 02115, USA} }
\date{\today}
\maketitle
%===================================================================
\begin{abstract}
We present a self-gravitating dyon solution of the
Einstein-Yang-Mills-Higgs equations of motion in
asymptotica\-lly AdS space. The back reaction of gauge and
Higgs fields on the space-time geometry leads to the metric of
an asymptotically AdS black hole. Using the gauge/gravity
correspondence we analyze relevant properties of the finite
temperature quantum field theory   defined on the boundary. In
particular we identify an order operator, characterize a phase
transition of the dual theory on the border and also compute
the expectation value of the finite temperature Wilson loop.
\end{abstract}
%===================================================================
\section{Introduction}
The AdS/CFT correspondence \cite{Maldacena}-\cite{Witten1998}
originally proposed  to connect gauge  and string theories has
recently  became a powerful tool  to study strong coupling
physics in systems of interest in quantum field theory and
condensed matter physics from the analysis of the low energy
approximation of string theories, namely gravitational theories
\cite{Gubser1}-\cite{HHH2} (for references on recent progress
see for example \cite{R1}).

In a previous work \cite{LMS2} we have used such gauge-gravity
duality  to  relate a $d=3+1$  Yang-Mills Higgs model in a
Schwarzschild-Anti-de Sitter black hole background with a
$d=2+1$ quantum field theory defined on the boundary. In this
way we were able to study  the strong coupling regime of the
theory on the boundary from the classical dyon solution found
in the bulk, showing that the $2+1$ system undergoes a second
order phase transition and exhibits the condensation of a
composite charge operator.

It is the purpose of the present work to extend our study to the
case in which there is back reaction on the space-time geometry so
that the complete Einstein-Yang-Mills-Higgs (EYMH) system of
equations of motion has to be solved, with the condition that the
metric solution corresponds to a black hole in asymptotically AdS
space.

Self-gravitating dyon solutions of the EYMH equations in
asymptotica\-lly AdS space,  regular both at the origin and at
infinity, were constructed in \cite{LS}-\cite{LMS}. It was
found that, as the value of the dimensionless parameter $G_N
H_0^2$ grows,  the metric function develops a minimum which
approaches to zero at a critical value ($G_N$ is the Newton
constant and $H_0$ the v.e.v. of the Higgs scalar). Above the
critical value a singularity at the origin develops indicating
the possibility of a threshold horizon. In fact, black hole
solutions that are finite at spatial infinity and at one
horizon have been shown to exist in systems like the one we are
interested for the case of asymptotically flat space
\cite{ewein1}-\cite{RT}. One of the purposes of the present
work is precisely to construct such black hole solutions in
asymptotically $AdS_4$ space, having finite mass and non-zero
horizon radius. Since the black hole naturally introduces a
temperature we shall be able to analyze relevant properties of
the finite temperature quantum field theory defined on the
boundary using the gauge/gravity correspondence.

It is interesting to notice that, in the context of the AdS/CFT
correspondence, gauge fields in the $AdS_{d}$ bulk can induce a
field theory on the  boundary containing dynamical gauge fields
for low dimensions  ($d \leq 4$) \cite{Aharony}-\cite{Maeda}.
It then becomes of interest to compute the expectation value of
the finite temperature Wilson loop of the theory at the
boundary which, according to   the gauge/gravity correspondence
\cite{Rey}-\cite{Brand}, is related to the Nambu-Goto action in
the bulk associated to the dyon solution.

~

The plan of the paper is the following. In section 2 we present
the model in the bulk $\mathbf{M}$ and propose a spherically
symmetric ansatz that reduces the EYMH equations of motion to a
a nonlinear system of coupled ordinary differential equations.
Such ansatz corresponds to a $\partial \mathbf{M}
=\mathbf{S}^2\times \mathbf{S}^1$  boundary (with the time
direction compactified to $ \mathbf{S}^1$). We also consider
the appropriate change of variables leading to $\partial
\mathbf{M} = \mathbf{R}^2\times \mathbf{S}^1$ boundary which
could be of interest in view of condensed matter applications.
In section 3 we discuss the appropriate conditions  on the
fields and metric in order to have a solution giving a black
hole in  asymptotically $AdS_4$ space and the gauge and Higgs
fields of a dyon. We construct numerically such a solution and
discuss its properties. We then apply the holographic
correspondence to analyze the behavior of the system defined on
the border as the temperature changes, identifying an order
parameter and characterizing the phase transition that the
$2+1$ system undergoes. Section 4 is devoted to the calculation
of the Wilson loop and binding energy for the theory in the
border. Finally, we present a summary and discussion of our
results in section 5.

\section{The model}

 As in ref.\cite{LMS2}, we consider a gravity-Yang-Mills-Higgs
system with gauge group $SU(2)$ and the scalar field in the
adjoint representation, in a 4 dimensional space-time with
Minkowski signature $(-,+,+,+)$. The action takes the form
\be S = S_G + S_{YM} + S_H = \int d^4x\,\sqrt{|g|}\; ( L_G + L_{YM}
+ L_H ) \label{1} \ee
with \be L_G = \frac{1}{2\,\kappa^2}  \left(  R - 2\,\Lambda \right)
\label{2} \ee
\be L_{YM} = -\frac{1}{4e{}^2}\; F_{\mu\nu}^a F^{a\,\mu\nu}
\label{3} \ee
\be L_H = -\frac{1}{2}    D_\mu H^a \; D^\mu H^a -
\frac{\lambda}{4}\; ( H^a H^a - H_0{}^2 )^2 \label{4} \ee
Here  $ \kappa^2\equiv 8\;\pi\; G_N$ with  $G_N$ the Newton
constant; $e$ the gauge coupling   and $\Lambda$   the
cosmological constant (we take $\Lambda < 0$ which corresponds
for our conventions and in the absence of matter, to anti-de
Sitter space). For simplicity we will consider the BPS limit
$\lambda/e^2 = 0$. The field strength $F^a_{\mu \nu}$
($a=1,2,3$) and  the covariant derivative $D_\mu$ acting on the
Higgs triplet $H^a$ are defined as
\be F^a_{\mu \nu} = \partial_\mu A_\nu^a - \partial_\nu A_\mu^a
+ \varepsilon^{abc}A_\mu^b A_\nu^c \;, \;\; \;\;\;
D_\mu H^a = \partial_\mu H^a + \varepsilon^{abc} A_\mu^b H^c
\label{6} \ee

The most general static spherically symmetric form for the
metric in $3$ spatial  dimensions together with the
t'Hooft-Polyakov-Julia-Zee ansatz for the gauge and Higgs
fields in the usual vector notation for internal degrees of
freedom reads \cite{LS}-\cite{LMS}
\ba ds^2 \!\!\!\! &=& - \mu(x)\; A(x)^2\; dt^2 + \mu(x)^{-1}\;
dr^2 + r^2\; d
\Omega_2\nonumber\\
\vec A &=& dt \; e\; H_0\; J(x)\; \vr - d\theta\; (1 - K(x) )\; \vfi
+ d\varphi \; (1 - K(x) )\; \sin\theta\; \vteta\nonumber\\
\vec H &=& H_0\; H(x)\; \vr \label{ansatz} \ea
where $H_0$ sets the mass scale ($[H_0] = m^1$) and we  have
introduced the dimensionless radial coordinate $\; x\equiv e\;
H_0\; r$.

\subsection*{From $\mathbf{S^2}$ to $\mathbf{R}^2$}

As stated in the introduction, we are interested in the
physical system defined on the boundary $\mathbf{\partial M}$
using the dual classical description of the gravity system
governed by the action (\ref{1}) in the 4 dimensional manifold
$\mathbf{M}$ which is, asymptotically, anti de Sitter space.
This means that after compactifying in the time direction, one
has $\mathbf{\partial M} = \mathbf{S^2} \times \mathbf{S^1}$.
Now, following  the approach in \cite{Witten2} one can trade
such boundary  with $\mathbf{\partial M} = \mathbf{R^2} \times
\mathbf{S^1}$ by introducing a dimensionless  parameter $R$
through a change coordinates $(t,x,\theta,\varphi)\rightarrow
(\tau, y , x^1,x^2)$ (or equivalently
$(t,x,\theta,\varphi)\rightarrow (\tau,y,\rho,\varphi)$), and
then taking the $R \to \infty$ limit. As we did in
ref.\cite{LMS2}, we consider the change of variables
\begin{align}
&\tau = R\; t\cr
&y = (\gamma_0/R)\;x = r/(LR) \cr
&x^1 = 2\;RL\;\tan (\theta/2)\; \cos\varphi = \rho\;
\cos\varphi\cr
&x^2 = 2\;RL\;\tan (\theta/2)\;\sin\varphi =
\rho\; \sin\varphi
\end{align}
together with the field redefinitions
\begin{align}
&\tilde A(y)= A(x)\;,  &&\tilde f(y) = R^{-2}\;\mu(x)\;, \cr
&\tilde H(y) = H(x)\;, &&\tilde K(y) = R^{-1}\;K(x) \;,  && \tilde J(y) = R^{-1}\;J(x)
\end{align}
After this change, the ansatz (\ref{ansatz}) becomes
\begin{align}
&ds^2 = - \tilde f(y)\; \tilde A(y)^2\; d \tau^2 +
L^2\;\frac{ dy^2}{\tilde f (y)} + \; y^2
\frac{{d x^1}^2  + {dx^2}^2}{(1+\frac{\rho^2}{4\,(LR)^2})^2}\cr
&\vec A = d\tau \; e\; H_0\; \tilde J(y)\;
\left(\frac{4\,LR\,\rho}{\rho^2 + 4\,(LR)^2}\;\check e_\rho +
\frac{4\,(LR)^2 - \rho^2}{\rho^2 + 4\,(LR)^2}\;\check e_3\right)\cr
&+ \frac{4\,LR^2}{\rho^2 + 4\,(LR)^2}\!\!\left(\frac{1}{R} -\tilde
K(y)\!\!\right)\!\!\! \left[ \left(\frac{4\,(LR)^2 - \rho^2}{\rho^2 +
4\,(LR)^2}\;\check e_\rho -
 \frac{4\,LR\,\rho}{\rho^2 + 4\,(LR)^2}\;\check e_3\right)\!\!
\rho\;d\varphi -\check e_\varphi\;d\rho\right]\cr
&\vec H = H_0\; \tilde H(y)\;
\left(\frac{4\,LR\,\rho}{\rho^2 + 4\,(LR)^2}\;\check e_\rho +
\frac{4\,(LR)^2 - \rho^2}{\rho^2 + 4\,(LR)^2}\;\check e_3\right)
\label{ansatz2}
\end{align}
and the equations of motion for the Einstein-Yang Mills-Higgs
system take the form
\begin{align}
&\left( y\; \tilde f(y)\right)' =\frac1{R^2} + 3 y^2 -\kappa{}^2\,H_0{}^2\;
\left( \tilde f (y)\; \tilde V_1 + \tilde V_2 +
\frac{y^2}{2}\;\frac{\tilde J'(y)^2}{\tilde A(y)^2} + \frac{\tilde
J(y)^2\,\tilde K(y)^2}{\tilde f(y)\,\tilde A(y)^2}\right)\cr
&y\;\tilde A'(y)  = \kappa{}^2\, H_0{}^2 \left( \tilde V_1 +
\frac{\tilde J(y)^2\,\tilde K(y)^2}{\tilde f(y)^2\,\tilde A(y)^2}
\right)\;\tilde A(y) \cr
&\left( \tilde f(y)\, \tilde A(y)\,\tilde K'(y) \right)'
= \tilde A(y)\; \tilde K(y)\; \left( \frac{\tilde K(y)^2
-\frac{1}{R^2}}{y^2} + \frac{1}{\gamma_0{}^2}\;\tilde H(y)^2 -
\frac{1}{\gamma_0{}^2}\;\frac{\tilde
J(y)^2}{\tilde f(y)\,\tilde A(y)^2} \right)\cr
&\left(
y^2\,\tilde f(y)\,\tilde A(y)\,\tilde H'(y) \right)'= 2\tilde
A(y)\;\tilde H(y)\;
\tilde K(y)^2 \cr
&\tilde f(y)\,\tilde A(y)\;\left( \frac{y^2\tilde J'(y)}{\tilde
A(y)} \right)' = 2\;\tilde J(y)\,\tilde K(y)^2
\label{eqmatt2}
\end{align}
where
\be\label{poten2} \tilde V_1 = \gamma_0{}^2\;\tilde K'(y)^2 +
\frac{y^2}{2}\;\tilde H'(y)^2 \; , \;\;\;\;\; \tilde V_2 =
\gamma_0{}^2\;\frac{ (\tilde K(y)^2 -\frac{1}{R^2})^2}{2\; y^2}
\nonumber \ee
Here we have defined the dimensionless parameter
\be \gamma_0^2 \equiv - \frac{\Lambda}{3 \,e{}^2\, H_0{}^2} =
\frac{1}{L^2 \,e{}^2\, H_0{}^2}\qquad,\qquad L^2 \equiv
-\frac{3}{\Lambda} \label{gamacero} \ee

In the $R \to \infty$ limit  ansatz (\ref{ansatz2}) becomes
\begin{align}
&ds^2 = - \tilde f(y)\; \tilde A(y)^2\; d \tau^2 +
L^2\;\frac{ dy^2}{\tilde f (y)} + \; y^2 \left({d x^1}^2  +
{dx^2}^2\right)\cr
&\vec A = d\tau \; e\; H_0\; \tilde J(y)\;\check e_3 +
\frac{\tilde K(y)}{L} (\check e_1 dx^1 - \check e_2 dx^2) \cr
&\vec H = H_0\; \tilde H(y)\; \check e_3 \label{ansatz22}
\end{align}
Then system (\ref{eqmatt2})  reduces to
\begin{align}
&\left( y\; \tilde f (y)\right)' = 3y^2
-\kappa^2\,H_0{}^2\; \left( \tilde f (y)\; \tilde V_1 +
\tilde V_2 + \frac{y^2}{2}\;\frac{\tilde J'(y)^2}{\tilde
A(y)^2} + \frac{\tilde J(y)^2\,\tilde K(y)^2}{\tilde
f(y)\,\tilde A(y)^2}\right)\cr
&y\; \tilde A'(y)  =
\kappa^2\, H_0{}^2 \left( \tilde V_1 + \frac{\tilde
J(y)^2\,\tilde K(y)^2}{\tilde f(y)^2\,\tilde A(y)^2}
\right)\;\tilde A(y)\cr
&\left( \tilde f(y)\, \tilde A(y)\,\tilde K'(y) \right)'
=
\tilde A(y)\; \tilde K(y)\; \left( \frac{\tilde K(y)^2}{y^2} +
 \frac{1}{\gamma_0{}^2}\;\tilde H(y)^2  -\frac{1}{\gamma_0{}^2}
 \;\frac{\tilde J(y)^2}{\tilde
f(y)\,\tilde A(y)^2} \right)\cr
&\left( y^2\,\tilde
f(y)\,\tilde A(y)\,\tilde H'(y) \right)'= 2\tilde
A(y)\;\tilde H(y)\;
 \tilde K(y)^2\cr
&\tilde f(y)\,\tilde A(y)\;\left( \frac{y^2\tilde
J'(y)}{\tilde A(y)} \right)' = 2\;\tilde J(y)\,\tilde
K(y)^2\label{eqmatt222}
\end{align}
where now
\begin{align}
\tilde V_2 &=
\gamma_0{}^2\;\frac{ \tilde K(y)^4}{2\; y^2}
\end{align}

\section{The black  hole-dyon solution}
In reference \cite{LMS} we constructed self-gravita\-ting dyon
solution in asymptotically $AdS_4$ space which were regular
both at the origin and infinity. An important feature noted in
that work was that  as the   dimensionless parameter $\kappa
H_0$ grows, the metric function $\tilde f$ develops a minimum
which, at a critical value $(\kappa H_0)^c$, corresponds to a
zero of $\tilde f$. For values of $\kappa H_0 \geq (\kappa
H_0)^c$ dyon solution which are regular at the origin ceases to
exist; however, as explained in \cite{ewein1}-\cite{ewein2} one
can hope that solutions that are finite at spacial infinite and
at one horizon can be found and this will be precisely the
issue we shall analyze  in this section. We shall construct
here such kind  of black holes dyon solutions for the
Yang-Mills-Higgs system in asymptotically $AdS_4$ showing that
they have finite mass, non-zero horizon radius and,
consequently, an associated Hawking temperature so that one can
study finite temperature effects in the quantum field theory
defined on the boundary.

\subsection{Boundary conditions and properties of the solution}

We shall look for a solution to (\ref{eqmatt2}) with  a horizon
at  nonzero radius $y = y_h$ that we can chose, without loss of
generality, to be $y_h = 1$. At the horizon we impose the
conditions:
\be \tilde f(1) =  0   \;\;\; \text{and}   \;\;\;\; A(y),
\tilde K(y),\,\tilde H(y),\, \tilde J(y)/(y-1) \;\;\;
\text{regular at $y= 1$}\label{origin} \ee
so that
\begin{align}
&\tilde f(y) = f_1(y-1) + {\cal O}[(y - 1)^2]
\nonumber\\
&\tilde A(y) = a_0 + a_1(y-1) + {\cal O}[(y - 1)^2]
\nonumber\\
&\tilde H(y) =  h_0 + h_1 (y - 1) +
{\cal O}[(y - 1)^2]\nonumber\\
&\tilde K(y) =  k_0 + k_1 (y - 1) +
{\cal O}[(y - 1)^2]\nonumber\\
&\tilde J(y) =  j_1 (y - 1) + {\cal O}[(y - 1)^2] \label{horj}
\end{align}

Concerning  $y \to \infty$,   the   asymptotic behavior should
be
\begin{align}
&\tilde  f(y) = y^2 + F_0 +\frac{F_{1}}y + \cdots
\nonumber\\
&\tilde A(y) = 1 + \cdots
\nonumber\\
&|\vec{\tilde H}(y)| =     H_0 {-}  \frac{H_1}{y^3} +
\frac{H_2}{y^5} +  \frac{H_3}{ y^{4 + 2\nu}} + \cdots \; ,
\;\;\;\; \nu \in\mathbb{R}   \label{Has}\\
&\tilde K(y) = \frac{K_1}{y^{\nu +1}} + \frac{K_3}{y^{\nu
+3}}\; , + \cdots
  \label{K}\\
&\tilde  J(y) = J_0 +
   \frac{J_1}{y} + \ldots
   \label{asym}
\end{align}

\noindent Such behavior is consistent with the $\tilde K$
equation of motion  whenever the following condition holds
\be \frac1{\gamma_0^2} =   \nu(\nu +1)\label{eq} \ee
Equation (\ref{eq}) has two solutions,
\be \nu_\pm = -\frac12
\pm \frac12\sqrt{1 + 4{  m_W}^2L^2}
\label{beinti} \ee where ${  m_W}  = e
H_0 $ is the gauge boson mass. Only the $\nu_+$ root gives an
acceptable asymptotic behavior for $\tilde K(x)$ so that one
ends with
\be \tilde K(x)  = \frac{K_1 }{y^{\nu_+ +1}}  =
\frac{K_1 }{y^{\frac12(1 +  \sqrt{1 + 4{m_w}L^2}\,)}} \;\;\;\;
\;\;\;\; \;\;\;\;
 {\rm as}~ x \to \infty
 \label{beinti2}
\ee
As in \cite{LMS2} we define the field strength associated with the
surviving $U(1)$ symmetry in the form
\be {\cal F}^{U(1)}_{\mu\nu} \equiv \frac{H^a}{H_0} F_{\mu\nu}^a \ee
with the magnetic and electric fields,  and the corresponding
charges, given by
\begin{align}
&B^i = \frac12
\frac{\varepsilon^{ijk}}{\sqrt{ g^{(3)}}}{\cal F}^{U(1)}_{jk} \; ,
\;\;\; E^i = {\cal F}^{U(1)i0} \label{B} \\
&Q_m = \int  d^3 x \sqrt{ g^{(3)}} \nabla^{(3)}_i B^i =
-\frac{4\pi}{e}
\label{Qm}\\
&Q_e = \int d^3 x  \sqrt{ g^{(3)}}  \nabla^{(3)}_i E^i \label{Qe}
\end{align}
Note that   the spherically symmetric solution has a quantized $n=-
1$ magnetic charge (in units of $4\pi/e$).

Concerning the black hole temperature, it is given by
\be {\cal T} =
\frac{1}{4\pi L} \tilde A(y_h) \tilde f'(y_h) \; . \label{tempe} \ee
Now, taking into account that we have   $y_h = 1$ and defining
the dimensionless quantity
\be
 T \equiv \frac{\cal T}{eH_0}
\ee
one has
\be
 T = \frac{\gamma_0}{4\pi} a_0 f_1
\label{forT}
\ee
where $a_0$ and $f_1$ are defined in (\ref{horj}). Since $a_0$ and $f_1$
depend both on $\kappa$ and  $\gamma_0$ we have that $ T =
 T(\kappa, \gamma_0)$. Note that $T(0,\gamma_0) = (3/4\pi) \gamma_0$.

\subsection{Numerical analysis of the solution}
We have solved system  (\ref{eqmatt2}) with boundary conditions
(\ref{origin})-(\ref{asym}) using the relaxation method. In
this method the differential equations are replaced by a system
of finite-difference equations on a one-dimensional grid with
$N$ points (typically $N$=500 in our case). Then, the solution
is determined by starting from an initial guess and improving
it iteratively (see ref.\cite{relaxation} for details).

The system \eqref{eqmatt2} is composed by two first-order
differential equations (for the metric functions $\tilde f(y)$
and $\tilde A(y)$) and three second-order ones (for the matter
fields $\tilde K(y)$, $\tilde J(y)$, and $\tilde H(y)$), so to
implement the integration algorithm we need eight boundary
conditions. At the horizon, $y_h=1$, we choose
\begin{align}
&\tilde f(1)=0 \,, \cr
&\tilde J(1)=0 \,, \cr
&\tilde f'(1)\, \tilde A(1)\, \tilde K'(1) = \tilde A(1)\; \tilde K(1)\; \left( \tilde
K(1)^2 -\frac{1}{R^2} + \frac{1}{\gamma_0{}^2}\;\tilde
H(1)^2\right)\,, \cr
& \tilde f'(1)\,\tilde A(1)\,\tilde H'(1) =
2\tilde A(1)\;\tilde H(1)\; \tilde K(1)^2
\label{num-bc}
\end{align}
The last two conditions are derived by simply evaluating the
differential equations for $\tilde K$ and $\tilde H$ at the
horizon.

For the boundary corresponding to $y\to \infty$ we choose for
the numerical calculation a finite size $L=10$, and check that
the solutions are stable under variations of $L$.  We impose at
this boundary the conditions
\begin{align}
& \tilde A  = 1\, , \cr
& \tilde H = 1\, , \cr
& \tilde K \propto y^{-\nu -1}\, , \cr
& \tilde J = J_0
\end{align}

Following this procedure, we have searched for regular
solutions for different values of parameters $\kappa$ and
$\gamma_0$. The corresponding solution depends on just one
boundary condition, namely $J_0$.

As is standard in the relaxation method, its effectiveness is
primarily determined by the election of the initial trial
functions, especially in this system that comprise eight
first-order differential equations. We started by solving the
system in the region of small $\gamma_0$ and $\kappa$  where
the algorithm is very stable and the election of trial
functions is simple. Once we have found solutions in this
region, we select them as trial functions for the equations
with larger constants. As we will discuss below, for certain
critical values of the parameters there are no non-trivial
solutions to the differential system. Near these regions the
algorithm becomes unstable and more mesh points and a careful
election of the trial functions is needed in order to obtain
solutions with the desired accuracy.

Our analysis allows to identify, three  regions in parameter
space $(\kappa, \gamma_o)$ with a completely different
behavior, as depicted in figure 1. Region I is the one where
the gauge and Higgs field solution correspond to a dyon
solution, i.e. a soliton with both electric and magnetic
charge. Concerning the metric, it corresponds to an
asymptotically AdS dyonic black hole. We shall discuss in
detail this solution at the end of this section.

%%%%%%%%%%%%%%%%%%%%%%%%%%%%%%%%%%%%%%%%%%%%%%%%%%%%%%%%%%%%%%%%%%%%
\begin{figure}
\epsfxsize=3.5 in
\begin{center}
\leavevmode
\epsffile{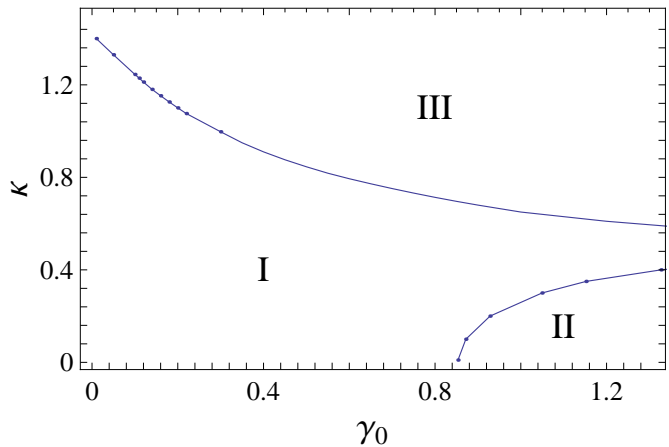}
\end{center}
\caption{The three regions in parameter space. Only
in region I the non trivial self-gravitating dyon solution
exists. Region II corresponds to the ``trivial'' solution and
no solution exists in region III whose frontier indicates the
appearance of a second horizon.}
\label{figure 1}
\end{figure}
%%%%%%%%%%%%%%%%%%%%%%%%%%%%%%%%%%%%%%%%%%%%%%%%%%%%%%%%%%%%%%%%%%%%

In region II, on the right,  only a ``trivial'' solution exists
for which the gauge and Higgs   fields take the form
\be \tilde H(y) = 1 \; , \;\;\; \tilde K(y) = 0 \; , \;\;\;
\tilde J = J_0(1-\frac1y) \label{40} \ee
Concerning the metric functions, they correspond to an AdS
dyonic black hole,  the result of the back reaction  on the
geometry
\be \tilde A(y) = 1 \; , \;\;\;  \tilde f(y) = \frac1{R^2} +  y^2 -
\frac{M}{y} + \frac{M-1 - 1/R^2}{y^2} \ee Here $M$ \be M= 1 +
\frac1{R^2} + \frac{\kappa^2 H_0^2 }{2}\left( J_0^2 +
\frac{\gamma_0^2}{R^4}\right) \ee
The reason for the existence of region II is well known: a too
large Newton constant destabilizes the dyon solution (its mass
grows and its radius decreases). This effect is enhanced in the
presence of a cosmological constant  so that both mechanisms
lead to the occurrence of a critical line after which the dyon
becomes gravitationally unstable and only the trivial solution
exists (see \cite{LMS2} and references therein for a more
detailed discussion).

Finally there existe a region III on the top of the figure,
which is bounded from below by a critical line indicating the
appearance of a second horizon. No solution, besides the
trivial one, exists above this line. For a more detailed
discussion on this issues see \cite{ewein1}-\cite{ewein2}).

We have analyzed the solutions both for  boundaries
$\mathbf{\partial M} = \mathbf{S^2} \times \mathbf{S^1}$ (fixed
$R$)  and for $\mathbf{\partial M} = \mathbf{R^2} \times
\mathbf{S^1}$ ($\lim R \to \infty$). Qualitatively the
solutions are very similar, thus for definiteness, we shall
present in detail the latter case.

Figures 2  shows a representative dyon solution arising in
region I. One can see that the Higgs fields rapidly attains its
symmetry breaking constant value while the magnetic and
electric fields (associated to $K$ and $J$ respectively)
concentrate in a spherical shell starting at the horizon.

%%%%%%%%%%%%%%%%%%%%%%%%%%%%%%%%%%%%%%%%%%%%%%%%%%%%%%%%%%%%%%%%%%%%
\begin{figure}[h]
\epsfxsize=3.5 in
\begin{center}
\leavevmode
\epsffile{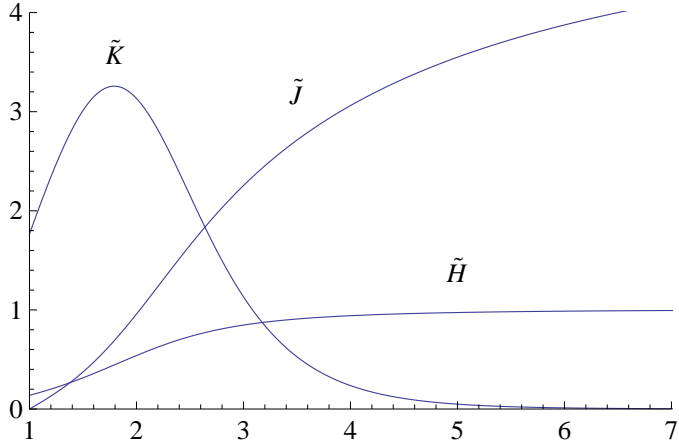}
\end{center}
\caption{Solution for the gauge field functions ($\tilde
K,\tilde J$ and scalar field $\tilde H$)  for $\gamma_0=0.1$,
$\kappa H_0= 0.5$ and $J_0 =5.5$. The solution exists starting
at the horizon $y=y_h=1$.}
\label{figure 2}
\end{figure}
%%%%%%%%%%%%%%%%%%%%%%%%%%%%%%%%%%%%%%%%%%%%%%%%%%%%%%%%%%%%%%%%%%%%
~

Concerning the metric function $\tilde f$, one can see that as
one increases $\kappa$ an outer minimum starts   to develop
(see figure 3) and moves downward, finally reaching zero and
becoming an extremal horizon at some critical value
$\kappa_{cr}$ a phenomenon already found for gravitational
monopoles in the case $\Lambda = 0$
\cite{ewein1}-\cite{ewein2}. This defines the critical line
between regions I and III. The coefficient $f_1$, the slope of
$\tilde f$ at the horizon has only a small variation with
$\kappa$, however the coefficient $a_0$ that represents the
function $\tilde A$ at the horizon, goes from 1, in the absence
of back reaction, to 0 at $\kappa=\kappa_c$.

~
%%%%%%%%%%%%%%%%%%%%%%%%%%%%%%%%%%%%%%%%%%%%%%%%%%%%%%%%%%%%%%%%%%%%
\begin{figure}[h]
\epsfxsize=5.5 in
\begin{center}
\leavevmode
\epsffile{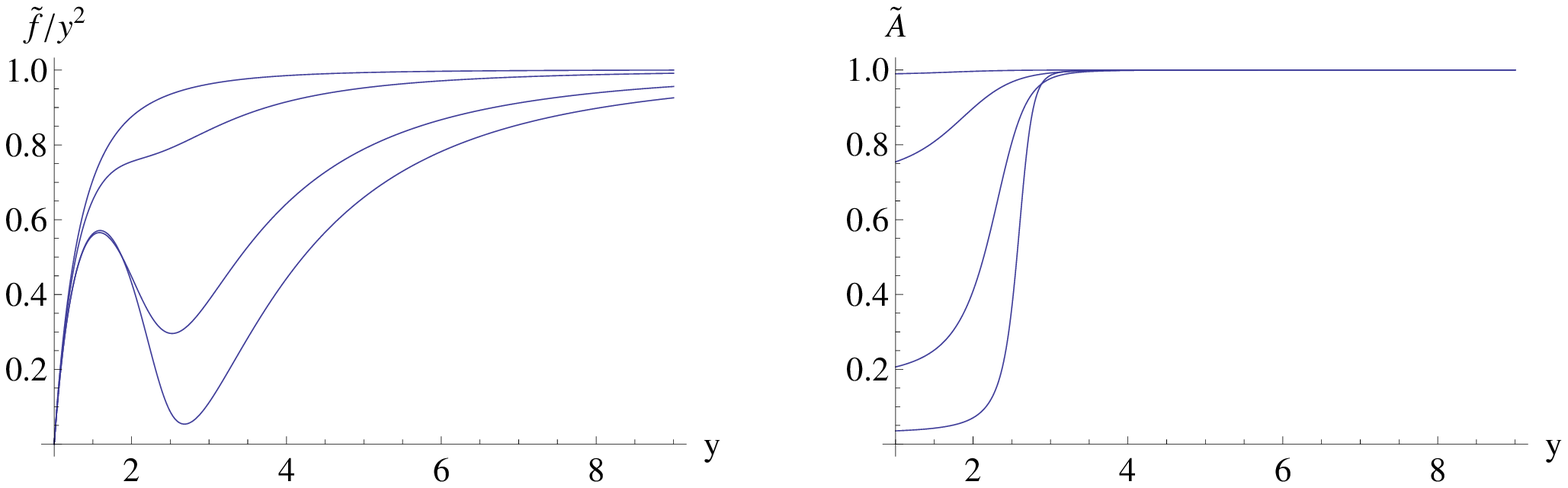}
\end{center}
\caption{Solution for  the metric functions $\tilde
f$ and $\tilde A$ as $\kappa H_0$ grows from $\kappa=0.1$
(highest curve) to $\kappa H_0= 1.2$ (lowest curve) with
$\gamma_0=0.1$ and $\tilde J_0=5.5$. For this case
$\kappa_{cr}=1.24$.}
\label{figure 3}
\end{figure}
%%%%%%%%%%%%%%%%%%%%%%%%%%%%%%%%%%%%%%%%%%%%%%%%%%%%%%%%%%%%%%%%%%%%

We conclude by noting that for small $\kappa$ the results
described above for the solution in region II are qualitatively
similar to those obtained in \cite{LMS2} the case in which back
reaction is ignored and a Schwarzchild-AdS$_4$ black hole is
taken as a background. This can be explained if one assumes the
existence of a weak gravity limit so that the $\kappa$
dependence is smooth and the solutions change continuously.
However, even in this situation similarities are lost for
certain ranges of parameters $(\kappa, \gamma_0)$, as the phase
diagram presented in Fig.1 exhibits  a critical line that
signals the existence of a second horizon, above which there
are no non-trivial solutions. One can associate a Hawking
temperature to this second horizon but since the bulk to be
considered within the gauge/gravity duality would correspond
just to the exterior of this second horizon, where only a
trivial solution exists, no relevant physics could arise in
this case.

\subsection{Holographic correspondence at finite temperature}
As explained in section 3.1, the Hawking temperature, as given
in eqs.(\ref{tempe})-(\ref{forT}), depends on the values of
parameters $\kappa$ and $\gamma_0$  , $T = T(\kappa,
\gamma_0)$. Once these parameters are chosen, the coefficients
$a_0$ and $f_1$ of the metric functions expansion near the
horizon can be calculated and then $T$ determined from
eq.(\ref{forT}).

According to the AdS/CFT correspondence \cite{Maldacena}
-\cite{Witten1998}, properties of the dual 3 dimensional field
theory defined on the boundary can be read from the  asymptotic
behavior of the solution of the system in the bulk. In this approach
the Hawking temperature $T(\kappa,\gamma_0)$ corresponds to the
temperature of the $d=3$ system on the boundary.

Let us first consider the vacuum expectation value in the $d=3$
field theory  for the   operator ${\cal O}_K$, dual to the
function $K$, associated with the magnetic field on the bulk.
It follows from the identification $\langle O_K \rangle \sim
K_1 $  with  $K_1$ defined in eq.(\ref{K}) that $K_1 = K_1(T)$
can be taken as an order parameter for the system in the
border. As discussed for different models
\cite{Gubser1}-\cite{LMS2} one can interpret this result by
stating that a condensate is formed  above a black hole horizon
because of a balance of gravitational and electrostatic forces.
One can from eqs.(\ref{beinti})-(\ref{beinti2})  see that the
dimension $\Delta[O_K]$  of the order operator is given by
\cite{Aharony}
\be \Delta [O_K] = \frac32 +\frac12\sqrt{1 + 4m_W^2L^2} = 2 +
\nu_+ \ee

 From the   solution that  we have found numerically we
conclude that a finite temperature continuous symmetry breaking
transition takes place so that the system condenses  at a
critical temperature $T_c$, as can be seen from the behavior of
$K_1(T)$ for $T \approx T_c$ in figure 4.

%%%%%%%%%%%%%%%%%%%%%%%%%%%%%%%%%%%%%%%%%%%%%%%%%%%%%%%%%%%%%%%%%%%%
\begin{figure}
\epsfxsize=3.5 in
\begin{center}
\leavevmode
\epsffile{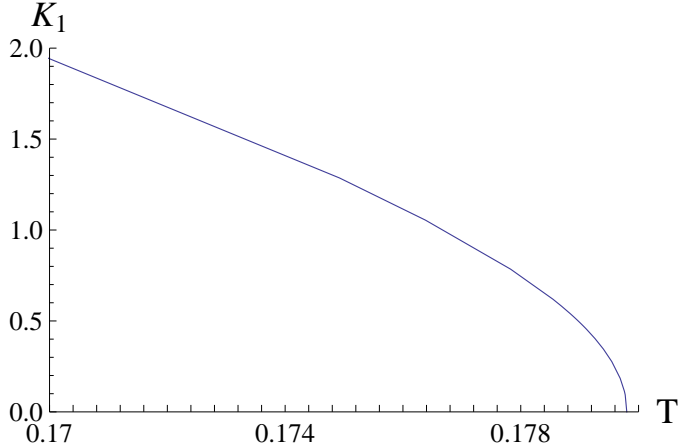}
\end{center}
\caption{The $K_1$ coefficient in the gauge field
$A_i$ asymptotic expansion as a function of temperature for
{$J_0 = 5.5$} and $\kappa H_0 =0.2$. To the right of the
critical temperature (uncondensed phase) the solution
corresponds to the trivial one, $K_1 = 0$. The critical
temperature for this choice of parameters is $T_c = 0.1798$.}
\label{figure 4}
\end{figure}
%%%%%%%%%%%%%%%%%%%%%%%%%%%%%%%%%%%%%%%%%%%%%%%%%%%%%%%%%%%%%%%%%%%%

~

By fitting the curve we see that near $T_c$ one has a typical
second order phase transition with power behavior of the form
\be \langle O_K \rangle \sim  K_1 \propto (T - T_c)^{1/2} \;
\;\;\; {\rm as} \; \;\;\; T \to T_c \label{transi} \ee
The free energy $\mathcal{F}$ associated to the dyon solution
can be calculated from the  the on-shell Euclidean action
action,
\be \mathcal{F} = T S_E|_{on~shell} \ee
We have numerically computed $\mathcal{F}$  for the dyon
solution and compared it with the free energy in the
uncondensed phase which corresponds to the trivial  solution
(\ref{40}). We plot in figure 5 the free energy density
difference $\Delta \mathcal{F}$, which can be seen to be
continuous at $T = T_c$.

%\vspace{0.5 cm}

%%%%%%%%%%%%%%%%%%%%%%%%%%%%%%%%%%%%%%%%%%%%%%%%%%%%%%%%%%%%%%%%%%%%
\begin{figure}
\epsfxsize=3.5 in
\begin{center}
\leavevmode
\epsffile{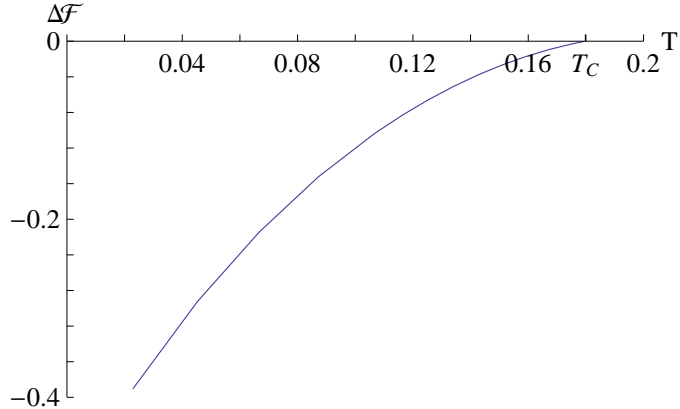}
\end{center}
\caption{The free energy difference between the condensed
and the uncondensed phases. The figure corresponds to $\kappa H_0 =
0.2$. To the left of the corresponding critical temperature
$\mathcal{F}_\text{dyon} < \mathcal{F}_\text{trivial}$.}
\label{figure 5}
\end{figure}
%%%%%%%%%%%%%%%%%%%%%%%%%%%%%%%%%%%%%%%%%%%%%%%%%%%%%%%%%%%%%%%%%%%%
~

%%%%%%%%%%%%%%%%%%%%
Other quantities of physical interest like the superconducting
charge density can be also computed following the same
approach. The main conclusion we can draw form the calculations
is that, concerning regions I and II in parameter space, the
results obtained by taking into account the back reaction of
the matter on the geometry are similar to those obtained in
ref.~\cite{LMS2} where the Yang-Mills-Higgs system is studied
in an AdS black hole background.

On the other hand, the main difference with the fixed
background case is that here we find a region in parameter
space where a second horizon arises and no non-trivial solution
exists. This implies that one cannot arbitrarily increase
Newton's constant and the cosmological constant and have a
nontrivial solution leading to a phase transition on the
border.

%%%%%%%%%%%%%%%%%%

It is important to mention that in the gauge-gravity duality
the extra dimension $r$ does not describe an additional
dimension but parameterizes the energy scale in the $2+1$ dual
field theory. So, the results described above indicate that the
theory in the boundary corresponds to a system undergoing a
superconductor phase transition as shown by the non-trivial
order parameter that reflects the existence of a condensate.

Let us recall at this point that, as explained before, the
$SU(2)$ gauge invariance of the Lagrangian in the bulk is
spontaneously broken to a $U(1)$ local symmetry due to the
non-trivial condition one imposes to the Higgs field at
infinity.  Such gauge symmetry in the bulk corresponds to a
$U(1)$ global symmetry in the dual field theory, broken by the
condensate which also breaks the rotation symmetry in the
spatial plane. A combination of the two, the diagonal subgroup,
is preserved. Concerning the fate of the corresponding
Goldstone boson, in a superconductivity context one should
start by coupling the theory on the boundary to a dynamical
photon. It can be seen that the Goldstone boson is eaten by the
photon, which becomes massive \cite{HHH2}.

As it is well known it is the $SU(2) \to U(1)$ symmetry
breaking pattern in a gauge theory coupled to a Higgs field
that allows the existence of regular monopoles and electrically
and magnetically charged dyons. One can think that the further
breaking of the local $U(1)$ to the global $U(1)$ in the
boundary could be at the origin of the
superfluidity/superconductivity phase regime in the presence of
the dyons. Electric repulsion between dyons and the charged
horizon compensates gravitational attraction and this is the
reason why the superconducting layer floats above the horizon.

Let us end this section by mentioning that it is possible to
perform a simple analytic treatment leading to qualitative
results in agreement with the detailed numerical analysis
discussed above. The method starts by changing the radial
coordinate to $z=r_h/r$ so that the horizon is located at $z=1$
and the boundary at infinity at $z=0$. One then proceeds to
expand functions $H, K$ and $J$ near $z=1$ and near $z=0$ and
matching them at $z=1/2$ by imposing continuity and smoothness.
Using this method, originally proposed in this context in
\cite{Gregory} for a simpler model, one can determine the
coefficients of both expansions and compute relevant physical
quantities. An analysis of this kind applied for a general
family of models is in progress and we can advance that one has
a good agreement in the case of the model discussed here.

\section{The Wilson loop}
As stated in the introduction, in  low dimensional cases ($d
\leq 3$), gauge fields in asymptotically AdS$_{d+1}$ spaces can
induce on the boundary a field theory with dynamical gauge
fields \cite{Wi2003}-\cite{Maeda}. On this basis, following the
gravity/gauge duality approach \cite{Rey}-\cite{MaldacenaW}, we
shall compute the Wilson loop for such gauge fields in $2+1$
dimensions in terms of the Nambu-Goto action for the string
world sheet defined from the metric solution found in the
precedent section. Since such solution corresponds to an
asymptotically AdS black hole, our calculation corresponds to
finite temperature \cite{ReyThe}-\cite{Brand}.

We start from the metric (\ref{ansatz2})
\be ds^2 = G_{tt} dt^2 +
G_{yy}dy^2 + L^2y^2 (d\theta^2 + \sin^2 \theta d\varphi^2) \ee
where
\be
G_{tt} = -A^2(y) \tilde f(y) \;, \;\;\;
G_{yy} = \frac1{\tilde f(y)}
\ee
and  consider a  string world-sheet   parameterized in the
static gauge according to
\begin{align}
&t(\tau,\sigma)= \tau\; , && \varphi(\tau,\sigma)=
\frac{\textit{l}\, \sigma} {L\sin\bar\theta}
\; , &&\theta(\tau,\sigma) =\bar\theta \; ,
 && y(\tau,\sigma) = y(\sigma)
\end{align}
where $\bar\theta$ is a constant and $y(\sigma)$ defines the
embedding, with $\sigma \in (-1/2,1/2)$. The quark $q$ and
anti-quark $\bar q$ are located at points $(\varphi=\alpha/2,
\theta = \bar \theta)$ and $(\varphi= \alpha/2, \theta = \bar
\theta)$ respectively, separated a distance $l$ along the
``parallel'' $\theta = \bar \theta$.

The resulting induced metric $h_{\alpha\beta}(\sigma)\equiv
G_{MN}(X)\,\partial_\alpha X^M(\sigma)\partial_\beta
X^N(\sigma)\;$ reads
\be h = G_{tt}\;(d\sigma^0)^2 + \left(
l^2\;y(\sigma)^2+  G_{yy}\;y'(\sigma)^2\right)\;(d\sigma)^2 \ee
so that the Nambu-Goto action $S_{NG}=T_s\;\int
\,d\sigma\;\sqrt{-\det h_{\alpha\beta}(\sigma) }$ can be
written as
\be S_{NG}= -T_s\;L\;\int_{-\frac{l}{2L}}^{+\frac{l}{2L}}\,dx\;
\sqrt{ F(u)^2 + G(u)^2\;u'(x)^2}\label{ng} \ee
with $u(x)\equiv y(L\, x/l)$, $F(u)\equiv u\,
A(u)\,\mu(u)^\frac{1}{2}$ and $G(u)\equiv A(u)$.

A minimal action configuration has energy
\be\label{E} \tilde E(l) = T_s\;L\;\left( F(u_m)\;\frac{l}{L} +
2\,\int_{u_m}^{u_\infty}\;du\;
G(u)\;\sqrt{1-\frac{F(u_m)^2}{F(u)^2}} \right)  \ee
where $u_m$ is the minimum value taken by $u$ and the distance
$l$ between the quark and the antiquark is given by
\be l=2
\int_{u_m}^{\infty} \frac{F(u_m)
G(u)}{\sqrt{F(u)^2(F(u)^2-F(u_m)^2}} du \label{eles}
\ee

Energy $E$ as given by eq.(\ref{E}) diverges linearly for
$u_\infty \rightarrow \infty$. However by subtracting the bare
mass $m_q$ of the two quarks, each one represented by long
strings extended along the $u$-direction and puncturing the
boundary $u=u_\infty$ at $x=\pm\frac{l}{2L}$,
\be 2\; m_q =
2\;T_s\;L\;\int_{u_h}^{u_\infty}\;du\; G(u) \ee
we get a finite
binding energy in the form
\begin{align}
\label{qbarqpot1}
&E(l) \equiv \tilde E(l) -2\,m_q =
T_s\;L\; \left( F(u_m)\;\frac{l}{L} - 2\,
K\left(\frac{l}{L}\right)\right)\cr
&K\left(\frac{l}{L}\right)= \int_{u_m}^{u_\infty}\;du\; G(u)\;
\left( 1-\sqrt{1-\frac{F(u_m)^2}{F(u)^2}}\right) +
\int_{u_h}^{u_m}\;du\;G(u)
\end{align}
where $u_h=1$  is the position of the horizon. We can re-write
the quark-antiquark potential in the form,
\be\label{qbarqpot2}
\frac{E(l)}{l} = T_s\;\left( F(u_m) -
2\;\frac{K\left(\frac{l}{L}\right)}{\frac{l}{L}}\right) \ee

The quark-antiquark binding energy $E = E(l,T)$ can be
calculated numerically  by eliminating $u_m$ between
eq.(\ref{eles}) and eq.(\ref{qbarqpot1}). Before doing this,
let us analyze the behavior of each one of these functions in
terms of $u_m$.

One can see in figure 6 that there exists a maximal distance
$l_{max}$ between the quark-anti-quark pair, a phenomenon
already found in the study of the finite temperature Wilson
line in the large $N$ limit of $U(N)$ ${\cal N} = 4$ Super
Yang-Mills theory in 4 dimensions \cite{ReyThe}-\cite{Brand}.

Concerning $E = E(u_m)$, the generic behavior in a wide range of
parameters values is the one depicted in figure 7. Note that $E = 0$
for a value $u_m^c$ satisfying $u_m^c> u_m^{max}$ leading to $l^c <
l^{max}$. Thanks to these relations,  valid in the whole
parameter range that we have studied, the possibility of
multi-valuation of the energy $E$ as exposed by the figure is
avoided: as $l$ grows from $l=0$ and reaches $l^c$, the energy of
the string configuration is the same as the energy of a pair of free
quark-antiquark. From there on the answer given by formula
(\ref{qbarqpot2}) does not correspond anymore to the lowest energy
which is in fact the free quark energy $E=0$.

%%%%%%%%%%%%%%%%%%%%%%%%%%%%%%%%%%%%%%%%%%%%%%%%%%%%%%%%%%%%%%%%%%%%
\begin{figure}
\epsfxsize=3.5 in
\begin{center}
\leavevmode
\epsffile{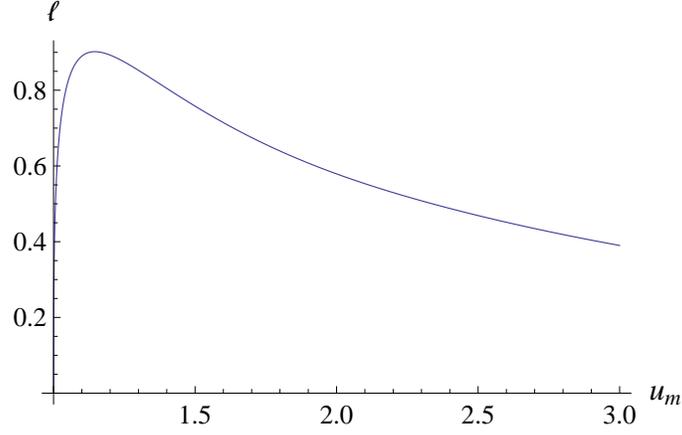}
\end{center}
\caption{The distance $l$ between the quark-antiquark pair as a
function of  $u_m$, the minimum value of the variable $u$. Note
that $l$ has a maximum possible value, $l^{max}$. The curve
corresponds to the choice $\gamma_0=0.1, \kappa H_0=0.5$ and
$J_0 =5.5$; for these parameters $u_m^{max} =   {1.15} $ and
$l^{max} = 0.90$ (in units of $eH_0$).}
\label{figure 6}
\end{figure}
%%%%%%%%%%%%%%%%%%%%%%%%%%%%%%%%%%%%%%%%%%%%%%%%%%%%%%%%%%%%%%%%%%%%

~

%%%%%%%%%%%%%%%%%%%%%%%%%%%%%%%%%%%%%%%%%%%%%%%%%%%%%%%%%%%%%%%%%%%%
\begin{figure}
\epsfxsize=3.5 in
\begin{center}
\leavevmode
\epsffile{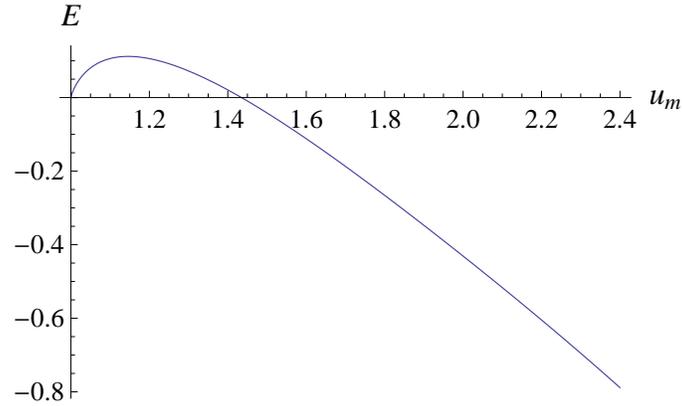}
\end{center}
\caption{The energy of the quark-antiquark pair relative to the free
quark configuration as a function of $u_m$, the minimum value
of the variable $u$ ($\gamma_0=0.1, \kappa H_0=0.5$ and $J_0
=5.5$). For these parameters one has (in units of $eH_0$,
$u_m^{c} =  {1.44} $ and $l^{c} = 0.79$.}
\label{figure 7}
\end{figure}
%%%%%%%%%%%%%%%%%%%%%%%%%%%%%%%%%%%%%%%%%%%%%%%%%%%%%%%%%%%%%%%%%%%%

Using eqs. (\ref{eles})-(\ref{qbarqpot2}) one can determine the
energy $E$ as a function of the distance $l$ between the quark
and the antiquark. The result is presented in figure 8 where we
see that there exists a critical distance $l^c$ at which the
quarks become free so that from there on one has a screening
behavior.

%%%%%%%%%%%%%%%%%%%%%%%%%%%%%%%%%%%%%%%%%%%%%%%%%%%%%%%%%%%%%%%%%%%%
\begin{figure}
\epsfxsize=3.5 in
\begin{center}
\leavevmode
\epsffile{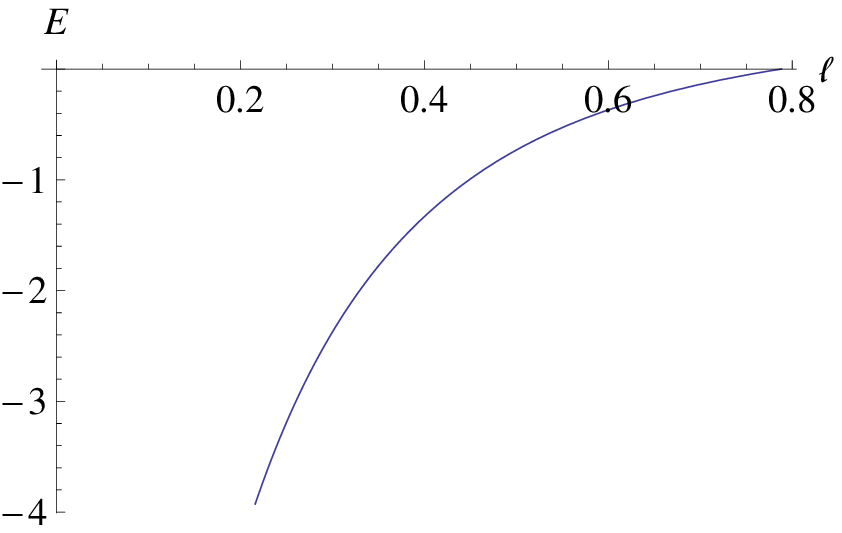}
\end{center}
\caption{The energy of the quark-antiquark pair as a
function of  $l$ ($\gamma_0=0.1, ~\kappa H_0=0.5$ and $J_0 =5.5$).
The graph corresponds to a temperature $T = 0.018$.}
\label{figure 8}
\end{figure}
%%%%%%%%%%%%%%%%%%%%%%%%%%%%%%%%%%%%%%%%%%%%%%%%%%%%%%%%%%%%%%%%%%%%

~

We have analyzed the behavior of $l^c$ as a function of the
temperature finding that such distance shortens as the
temperature grows, a behavior that can be taken as a signal
that the screening phenomenon is produced by the thermal bath.
A representative graph is given in figure 9.

\vspace{0.5 cm}

%%%%%%%%%%%%%%%%%%%%%%%%%%%%%%%%%%%%%%%%%%%%%%%%%%%%%%%%%%%%%%%%%%%%
\begin{figure}
\epsfxsize=3.5 in
\begin{center}
\leavevmode
\epsffile{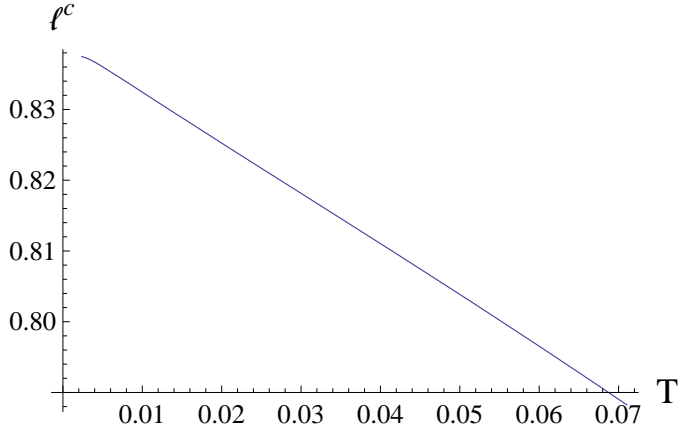}
\end{center}
\caption{The distance $l^c$ at which the quark and
the anti-quark become free as a function of temperature.
($\gamma_0=0.3$).}
\label{figure 9}
\end{figure}
%%%%%%%%%%%%%%%%%%%%%%%%%%%%%%%%%%%%%%%%%%%%%%%%%%%%%%%%%%%%%%%%%%%%

\section{Summary and discussion}
One of the purposes of the present work was to study the
$d=3+1$ dimensional Einstein-Yang-Mills-Higgs model and find
classical solutions to its equations of motion in
asymptotically AdS space. As it happens for asymptotically flat
space \cite{ewein1}-\cite{ewein2} we have shown  that one can
construct topologically stable self-gravitating dyons with a
metric resulting from the back reaction on the space-time
geometry solution that corresponds to a black hole in
asymptotically AdS space. We find such solutions in a certain
domain of parameter space which corresponds to region I in
figure 1. There is also a region where only the trivial
solution, for which the Higgs field takes its v.e.v. in the
whole space, the gauge field is a pure gauge  and the metric
corresponds to a Schwarzschild-AdS black hole (region II). As
for the behavior of the Higgs and gauge fields spherically
symmetric solutions, they are  very similar as those we
presented in ref. \cite{LMS2} for the case of a
Schwarzschild-AdS black hole background (no back reaction),
with the horizon radius inside the dyon core and the Higgs
field rapidly taking its vacuum expectation value. Finally
there is a critical line which marks the appearance of a double
horizon solution like the Reissner-Nordstrom metric. Above such
line, no solution exists.

Using the gauge-gravity correspondence, we were able to study
properties of the dual $d=2+1$ quantum field theory on the
boundary of the $3+1$ asymptotical AdS space where our
classical solutions were found. Because those solutions
correspond to a black hole metric, they have an associated
Hawking temperature $T = T(\gamma_0,\kappa)$ with parameters
$\gamma_0$ and $\kappa$ related to the cosmological and Newton
constants respectively. Then, varying those parameters in the
bulk classical solutions we were able to study the finite
temperature behavior of the vacuum expectation values of
operators defined on the boundary, identifying an order
parameter and showing that a second order phase transition
takes place.

%%%%%%%%%%%%%%%%%%%%%%%%%%%%%%%%%%%%%%%%%%%%%%%%%%%%%%%%%%%%%%
Finally, because we have taken into account the back reaction
of matter on geometry we were able to use the resulting black
hole metric to calculate the Nambu-Goto action for our black
hole solution related through the holographic correspondence to
the Wilson loop at finite temperature.
%%%%%%%%%%%%%%%%%%%%%%%%%%%%%%%%%%%%%%%%%%%%%%%%%%%%%%%%%%%%%%%
In this way we were able to evaluate  the binding energy $E =
E(l,T)$ of two external charges in the boundary (a
quark-antiquark pair). We found a {inverse power law behavior}
for $E$ as a function of the distance between quarks, up to a
maximal distance $l_c$ where $E$ vanishes indicating that the
quarks become free in a typical screening process which can be
attributed to the thermal bath.  It should be remarked that
this scenario is qualitatively the same already found on the
pioneering works on finite temperature Wilson loops in the
large $N$ limit of $U(N)$ ${\cal N} = 4$ supersymmetric
Yang-Mills theory in 4 dimensions \cite{ReyThe}-\cite{Brand}.

Let us note at this point that, following the approach applied
here, one can establish the existence of confinement for
Yang-Mills theory in three dimensions by computing Wilson loops
at finite temperature using the string/gauge correspondence
(see (\cite{Sonn}) for a detailed calculation).
%We have found,
%in contrast, a screening behavior so that one can discard the
%possibility that the model in the boundary is related to a
%gauge theory of the Yang-Mills type. One could expect that
%going beyond the classical approximation on the gravity side a
%more detailed description of the behavior in the boundary would
%allow to identify the precise dual field theory one is dealing
%with.
%%%%%%%%%%%%%%%%%%%%%%%%%%%%%%%%%%%%%%%%%%%%%%%%%%%%%%%%%%%%
We have found, in contrast, a screening behavior so that we can
discard the possibility that the model in the boundary could be
related directly to $2+1$ Yang-Mills. It is however well known
that when massive fermions are added to a $2+1$ gauge theory
the effective action resulting from integration of fermions
induces a Chern-Simons term and, as a result, confinement is
destroyed (see for example \cite{FS}). Then,  one can speculate
that the theory on the boundary could include a Chern-Simons
term. This idea may be tested by going beyond the classical
approximation on the gravity side which would require to embed
the gravity model into string theory. We hope to present a
thorough study on these issues in the future.
%%%%%%%%%%%%%%%%%%%%%%%%%%%%%%%%%%%%%%%%%%%%%%%%%%%%%%%%%%%%%
%We hope to present a thorough study on this issue in the
%future.

\vspace{1 cm}

\noindent\underline{Acknowledgments}  We would like to thank
Carlos N\'u\~nez and Jorge Russo for discussions and helpful
suggestions. This work was partially supported by
PIP1787-CONICET,  PICT20204-ANPCYT grants and by CIC and UNLP,
Argentina.

%%%%%%%%%%%%%%%%%%%%%%%%%%%%%%%%%%%%%%%%%%%%%%%%%%%%%%%%%%%%%%%

%%%%%%%%%%%%%%%%%%%%%%%%%%%%%%%%%%%%%%%%%%%%%%%%%%%%%%%%%%%%%%%%%%%%

\end{document}